\documentclass[10pt,a4paper]{article}
\usepackage{amsmath,amssymb,latexsym}
\usepackage{mathrsfs,graphicx}

\def\AFOUR{%
\setlength{\textheight}{8.5in}%
\setlength{\textwidth}{5.75in}%
\setlength{\topmargin}{-0.375in}%
\hoffset=-.5in%
\renewcommand{\baselinestretch}{1.17}%
\setlength{\parskip}{6pt plus 2pt}%
}

\AFOUR                                           

\expandafter\ifx\csname amssym.def\endcsname\relax \else\endinput\fi
\expandafter\edef\csname amssym.def\endcsname{%
       \catcode`\noexpand\@=\the\catcode`\@\space}
\catcode`\@=11
\def\undefine#1{\let#1\undefined}
\def\newsymbol#1#2#3#4#5{\let\next@\relax
 \ifnum#2=\@ne\let\next@\msafam@\else
 \ifnum#2=\tw@\let\next@\msbfam@\fi\fi
 \mathchardef#1="#3\next@#4#5}
\def\mathhexbox@#1#2#3{\relax
 \ifmmode\mathpalette{}{\m@th\mathchar"#1#2#3}%
 \else\leavevmode\hbox{$\m@th\mathchar"#1#2#3$}\fi}
\def\hexnumber@#1{\ifcase#1 0\or 1\or 2\or 3\or 4\or 5\or 6\or 7\or 8\or
 9\or A\or B\or C\or D\or E\or F\fi}

\font\tenmsa=msam10
\font\sevenmsa=msam7
\font\fivemsa=msam5
\newfam\msafam
\textfont\msafam=\tenmsa
\scriptfont\msafam=\sevenmsa
\scriptscriptfont\msafam=\fivemsa
\edef\msafam@{\hexnumber@\msafam}
\mathchardef\dabar@"0\msafam@39
\def\dashrightarrow{\mathrel{\dabar@\dabar@\mathchar"0\msafam@4B}}
\def\dashleftarrow{\mathrel{\mathchar"0\msafam@4C\dabar@\dabar@}}

\def\ulcorner{\delimiter"4\msafam@70\msafam@70 }
\def\urcorner{\delimiter"5\msafam@71\msafam@71 }
\def\llcorner{\delimiter"4\msafam@78\msafam@78 }
\def\lrcorner{\delimiter"5\msafam@79\msafam@79 }
\def\yen{{\mathhexbox@\msafam@55}}
\def\checkmark{{\mathhexbox@\msafam@58}}
\def\circledR{{\mathhexbox@\msafam@72}}
\def\maltese{{\mathhexbox@\msafam@7A}}
\def\circledS{{\mathhexbox@\msafam@73}}

\font\tenmsb=msbm10
\font\sevenmsb=msbm7
\font\fivemsb=msbm5
\newfam\msbfam
\textfont\msbfam=\tenmsb
\scriptfont\msbfam=\sevenmsb
\scriptscriptfont\msbfam=\fivemsb
\edef\msbfam@{\hexnumber@\msbfam}
\def\Bbb#1{{\fam\msbfam\relax#1}}
\def\widehat#1{\setbox\z@\hbox{$\m@th#1$}%
 \ifdim\wd\z@>\tw@ em\mathaccent"0\msbfam@5B{#1}%
 \else\mathaccent"0362{#1}\fi}
\def\widetilde#1{\setbox\z@\hbox{$\m@th#1$}%
 \ifdim\wd\z@>\tw@ em\mathaccent"0\msbfam@5D{#1}%
 \else\mathaccent"0365{#1}\fi}
\font\teneufm=eufm10
\font\seveneufm=eufm7
\font\fiveeufm=eufm5
\newfam\eufmfam
\textfont\eufmfam=\teneufm
\scriptfont\eufmfam=\seveneufm
\scriptscriptfont\eufmfam=\fiveeufm
\def\frak#1{{\fam\eufmfam\relax#1}}

\csname amssym.def\endcsname

\makeatletter
\def\section{\@startsection {section}{1}{\z@}{-3.5ex plus -1ex minus
 -.2ex}{2.3ex plus .2ex}{\large\sc}}
\def\subsection{\@startsection{subsection}{2}{\z@}{-3.25ex plus -1ex minus
 -.2ex}{1.5ex plus .2ex}{\normalsize\sc}}
\makeatother

\makeatletter
\@addtoreset{equation}{section}

\makeatother

\newcommand{\nc}{\newcommand}
\newcommand{\rnc}{\renewcommand}

\nc{\chap}[1]{{\clearpage}%
\begin{center}%
{\noindent\underline{\large\sc #1}}{\addcontentsline{toc}{section}{#1}}%
\end{center}%
{\vspace*{0.3cm}}}

\nc{\subs}[1]{{\vspace*{0.2cm}}%
{\noindent\underline{\small\sc
#1}}%
{\vspace*{0.2cm}}}

\nc{\be}{\begin{equation}}
\nc{\ee}{\end{equation}}
\nc{\bea}{\begin{eqnarray}}
\nc{\eea}{\end{eqnarray}}

\nc{\trac}[2]{{\textstyle\frac{#1}{#2}}}

\nc{\ex}[1]{\mbox{e}^{\,\textstyle#1}}

\nc{\CC}{\Bbb{C}}
\nc{\HH}{\Bbb{H}}
\nc{\PP}{\Bbb{P}}
\nc{\RR}{\Bbb{R}}
\nc{\ZZ}{\Bbb{Z}}
\nc{\II}{\Bbb{I}}
\nc{\EE}{\Bbb{E}}
\nc{\TT}{\Bbb{T}}
\nc{\DD}{\mathrm{I}\!\mathrm{D}}

\rnc{\d}{\delta}
\nc{\eps}{\epsilon}
\nc{\om}{\omega}

\nc{\symx}{\circledS}
\newsymbol\smallsmile 1360
\newsymbol\smallfrown 1361
\nc{\ad}{\mathop{\mbox{ad}}\nolimits}
\nc{\tr}{\mathop{\mbox{tr}}\nolimits}
\nc{\Tr}{\mathop{\mbox{Tr}}\nolimits}
\nc{\Det}{\mathop{\mbox{Det}}\nolimits}
\rnc{\det}{\mathop{\mbox{det}}\nolimits}
\nc{\rk}{\mathop{\mbox{rk}}\nolimits}
\nc{\del}{\partial}
\nc{\diag}{\mathop{\mbox{diag}}\nolimits}
\nc{\ra}{\rightarrow}
\nc{\Ra}{\Rightarrow}
\nc{\LRa}{\Leftrightarrow}
\nc{\lra}{\leftrightarrow}
\nc{\ot}{\otimes}
\rnc{\ss}{\subset}
\nc{\nul}{\noindent\underline}
\nc{\non}{\nonumber\\}
\nc{\mat}[4]{\left(\begin{array}{cc}#1&#2\\#3&#4\end{array}\right)}
\rnc{\lg}{\frak{g}}
\nc{\G}[3]{\Gamma^{#1}_{\;{#2}{#3}}}
\nc{\nam}{\nabla_{\mu}}
\nc{\nan}{\nabla_{\nu}}
\nc{\dx}{\dot{x}}
\nc{\tx}{\tilde{x}}
\nc{\dtx}{\dot{\tilde{x}}}
\nc{\te}{\tilde{e}}
\nc{\dte}{\dot{\tilde{e}}}
\nc{\dxl}{\dot{x}^{\la}}
\nc{\dxm}{\dot{x}^{\mu}}
\nc{\dxn}{\dot{x}^{\nu}}
\nc{\ddx}{\ddot{x}}
\nc{\ddxm}{\ddot{x}^{\mu}}
\nc{\ddxn}{\ddot{x}^{\nu}}
\nc{\dxi}{\dot{\xi}}
\nc{\ddxi}{\ddot{\xi}}
\nc{\lsf}{\ell_s^{\mathrm{eff}}}
\nc{\lpf}{\ell_p^{\mathrm{eff}}}
\nc{\sqg}{\sqrt{g^{11}}}

\nc{\bpm}{\begin{pmatrix}}
\nc{\epm}{\end{pmatrix}}

\nc{\der}{\mathrm{d}}

\begin{document}

\begin{center}
{\Large\sc Aspects of Plane Wave (Matrix) String Dynamics}
\end{center}
\vspace{0.5cm}

\begin{center}
{\large\sc Matthias Blau$\;{}^a$,
Martin O'Loughlin$\;{}^b$, Lorenzo Seri$\;{}^c$}\\[.8cm]
{\it ${}^a$ Albert Einstein Center for Fundamental Physics,
Institute for Theoretical Physics\\ Bern University,
Sidlerstrass 5, 3012 Bern, Switzerland }\\[.3cm]
{\it ${}^b$ University of Nova Gorica, Vipavska 13, 5000 Nova Gorica,
Slovenia}\\[.3cm]
{\it ${}^c$ SISSA, Trieste, Italy}
\end{center}

\vspace{.5cm}

We analyse two issues that arise in the context of (matrix) string
theories in plane wave backgrounds, namely (1) the use of Brinkmann-
versus Rosen-variables in the quantum theory for general plane waves
(which we settle conclusively in favour of Brinkmann variables), and (2)
the regularisation of the quantum dynamics for a certain class of singular
plane waves (discussing the benefits and limitations of regularisations
of the plane-wave metric itself).

\newpage

\section{Introduction}

As exact and potentially exactly solvable string backgrounds, 
plane waves have been extensively studied in string theory
(see e.g.\ \cite{at1} for a review of the early literature). 
As such they also provide an ideal setting for studying the 
issue and fate of time-dependent backgrounds and space-time singularities in string 
theory \cite{hs,Sanchez}. 
In the usual Brinkmann coordinates, a  plane wave metric has the form
\[ds^2= 2dz^+dz^- + A_{ab}(z^+)z^az^b (dz^+)^2 + \d_{ab}dz^a dz^b\;\;,\]
and this metric is singular iff the profile $A_{ab}(z^+)$ (= minus the matrix of 
frequency-squares for transverse string modes in the lightcone gauge $z^+\sim t$) 
diverges for some (finite) value of $z^+$.\footnote{What is meant by a singularity of the 
metric in 
the present context (where all scalar curvature invariants of the metric are
zero) is divergent tidal forces, manifested by a divergence of components of
the Riemann tensor in a parallel propagated frame (p.p.-singularities in the 
terminology of \cite{HE}). Thus plane waves with a singular profile
have null p.p.-singularities.} 
Among all singular plane waves, from the point of view of studying
string theory in singular backgrounds, those with a profile of the form
$A_{ab}(z^+)\sim (z^+)^{-2}$ are singled out by at least two facts:

\begin{enumerate}
\item Precisely these backgrounds were shown in \cite{bbop1,bbop2,gpsi}
to arise universally as the Penrose limits \cite{penrose,gueven,bfp}
of a large general class of singularities (of power-law type \cite{SI})
which encompasses all the standard string theory black hole and black
brane singularities as well as those of all (non-oscillatory/chaotic)
standard cosmological models. Thus these special singular plane waves
are not just toy-models of space-time singularities but honest 1st-order
approximations (in the spirit of the Penrose-Fermi expansion \cite{fermi}
around the Penrose limit metric) to realistic types of singularities.
\item These singular plane waves possess precisely those symmetries
(isometries and homotheties), in particular the boost/scale symmetry
$z^\pm \ra \lambda^\pm z^\pm$, that are required \cite{mmpwbb}
to implement the usual (flat space) Seiberg-Sen(-CSV)
\cite{seiberg,sen,sen2,csv} DLCQ derivation of matrix string theory
\cite{matrixstrings}
for these curved backgrounds, so that in principle one has available
a non-perturbative definition of string theory in these backgrounds
(the plane wave matrix big bang models of \cite{mmpwbb}).
\end{enumerate}

The above-mentioned boost/scale symmetry renders the
metric homogeneous away from the locus of the singularity at $z^+=0$
(see e.g.\ \cite{bfp,prt,mmhom}). Thus these metrics can be considered
as singular counterparts of the much studied (symmetric) plane waves with constant
profile $A_{ab}$ (which are homogeneous because of the invariance under
lightcone time translations), 
and this scale-invariance is reflected in (and impacts upon)
the quantisation of strings and particles in such a background
\cite{prt,mmhom,bopt,bbo}. In the lightcone gauge $z^+=t$, the string mode 
equations are simply harmonic oscillator equations with time-dependent 
frequencies $\omega_n(t)^2 = \omega(t)^2  + n^2$, with $\omega(t)^2\sim t^{-2}$
arising from the plane wave profile.
In particular, perturbative string theory
in these singular homogeneous plane waves (with weak string coupling
at the singularity) was analysed in detail in \cite{prt}, with the
conclusion that, although there is divergent particle production on the
world-sheet as one approaches the singularity at $t=0$ (cf.\ also
\cite{hs} and \cite{hs2}), there is a possibility that an analytic continuation of
the string wave-functions could lead to non-singular evolution through
the singularity. 

Curiously exactly the same (decoupled, Abelian) string mode equations
arise from the analysis of strong string coupling singularities in the
context of the (non-Abelian) matrix string models \cite{mmpwbb} for
singular plane waves, in which the strong string coupling implies weak
(time-dependent) Yang-Mills gauge coupling.  This comes about because,
as shown in \cite{mols}, near the singularity the usual
quartic interaction term for the non-Abelian matrix string coordinates
becomes irrelevant.\footnote{This is a stronger statement than just the fact
that the Yang-Mills coupling constant goes to zero, its validity relying \cite{mols} on the
precise time-dependence of the dilaton arising from the requirement
that the metric-dilaton configuration be a solution to the string
background equations of motion \cite{mmpwbb}.} Thus the near-singularity behaviour
of the matrix string coordinates is governed only by the decoupled
(non-self-interacting) kinetic and harmonic oscillator (mass) terms
$\sim \omega_n(t)^2$, as above, and one formally encounters
the same problems and issues as in the perturbative string theory model of \cite{prt},
now however at strong string coupling.

The (thus necessary) issue of regularisation of the particle and string dynamics in
these singular backgrounds was taken up again in \cite{craps1,craps2}.\footnote{See also the review 
\cite{craps4} for a discussion of the state of affairs.}
In particular, in \cite{craps2} a specific proposal was made to regularise the dynamics, 
namely by an explicit single-scale regularisation 
of the oscillator frequencies (equivalently, of the plane wave metric) 
of the form
\[\omega^2(t) \ra 
\Omega_{\epsilon}^2(t) = \epsilon^{-2}\Omega^2(t/\epsilon)
\]
in which the scale invariance of the original background is as much as possible
respected. In \cite{craps2,oetn} some minimal conditions
were proposed for such a regularisation to lead to a well-defined quantum evolution 
in the limit $\epsilon\ra 0$ that the regulator is removed. 
One of the purposes of this article is to analyse and scrutinise this
proposal in some more detail.

Before embarking on this, however, there is another issue that needs to be
addressed and settled once and for all which is a minor annoyance but which occasionally rears its ugly
head. This is the question whether one should analyse or quantise a plane
wave system in the Brinkmann coordinates used above, or whether one can
(either alternatively or even profitably) study the system in variables
based on Rosen coordinates, in which a plane wave metric takes the form
\[ ds^2 = 2 du dv + g_{ij}(u) dx^i dx^j\;\;.\]
In
particular, in the present context, metrics of this form appear to provide
attractive models of \textit{null cosmologies\/} \cite{prt}. However,
as is well known (and as we will recall at the beginning of section 2),
these Rosen coordinate systems are fraught with ambiguities, almost
invariably exhibiting spurious coordinate singularities, while the
Brinkmann coordinate system encodes directly geometrically invariant
and physical information about the space-time (coordinates $z^a$ are measures of proper
distance, and the profile $A_{ab}$ assembles the parallel orthonormal
frame components of the Riemann tensor). 

Since it has, in spite of this, occasionally been suggested (see e.g.\
\cite{narayan,narayan2}) that the Rosen system may provide
a better, because seemingly less singular, description of the quantum
dynamics, in section 2 we first quickly go through the (essentially undergraduate)
exercise of constructing and solving the quantum theory of Brinkmann and
Rosen particles and the (unitary) relation between them. We then revisit
the Brinkmann vs Rosen issue in this quantum context, stress that this is
not a question of Rosen vs Brinkmann quantum states but rather a matter of which 
operators one attributes physical significance to, and show that the plane wave
space-time origin of these models uniquely leads to the Brinkmann picture. 
This conclusion is then significantly strengthened and sharpened 
in section 2.4, where
we show that some rather remarkable and striking recent uniqueness results 
\cite{cortez1,cortez2,cortez3,cortez4, cortez5}
regarding the quantisation of scalar fields with time-dependent mass terms
imply that for strings in arbitrary plane wave backgrounds (1)
in the Brinkmann parametrisation there is a unique Fock quantisation
respecting the translation invariance along the worldsheet $S^1$
(namely the standard Fock space of a free massless scalar field)
such that the dynamics is unitarily implemented; and (2)
that there is no $S^1$-invariant Fock quantisation of the Rosen system
with unitary implementable time evolution
(in the singular case, with a mass term $\sim t^{-2}$, these results
hold for any $t \neq 0$). 

The upshot of this discussion is that the quantisation of (matrix)
strings in plane wave backgrounds needs to be performed in Brinkmann
variables (something that we had already advocated in \cite{mmpwbb} on
the related grounds that it is only in these variables that the scalars
have a canonical kinetic term), and that in spite of the time-dependence
of the mass-term / Hamiltonian
(which usually leads to quantisation ambiguities and phenomena like
particle production etc.) there is a unique Fock quantisation with unitary
time-evolution.

Having disposed of the Rosen representation, we then (re-)turn to the
issue of regularisation of the dynamics in singular scale-invariant plane
waves and the proposal of \cite{craps2}. We first specialise the results
of section 2 regarding expectation values to the singular plane waves
to illustrate how the plane wave singularity manifests itself in the
quantum theory. We then exhibit a family of single-scale regularisations
that satisfy the necessary conditions of \cite{craps2,oetn}, essentially
by a method that amounts to directly regularising the Rosen metric
rather than the Brinkmann wave profile / frequency. We analyse
the extent to which such regularisations can be considered to give a
non-singular evolution through $t=0$ in the limit that the regularisation
parameter $\eps\ra 0$, with the conclusion that no regularisation of
the single-scale form can provide this. More complicated regularisations
that achieve this do exist, however, and we comment on this as well as on
issues related to the regularisation of the dilaton. We end with a 
discussion of the results, in particular their implications for
matrix string models of plane wave null singularities, and their interpretation 
in terms of the Penrose limit origin of these space-times.

\section{Quantum Brinkmann vs Rosen Plane Wave Dynamics}

In this section we will analyse the relation between the quantum dyanmics
of particles and strings in the Brinkmann and Rosen parametrisations.
To set the stage and motivate the discussion of this section, we begin
by recalling the well-known ambiguities and spurious singularities afflicting the
Rosen description of the classical particle dynamics. 

\subsection{Classical Dynamics and Rosen Singularities}

In Brinkmann (resp.\ Rosen) coordinates, plane wave metrics have the form
\be
\begin{aligned}
ds^2&= 2dz^+dz^- + A_{ab}(z^+)z^az^b (dz^+)^2 + \d_{ab}dz^a dz^b\\
&= 2 du dv + g_{ij}(u) dx^i dx^j
\end{aligned}
\ee
($u=z^+$, the remaining coordinate transformation between the unique Brinkman form of
the metric and the non-unique Rosen form is given in \eqref{rcbc2}). 
For present purposes nothing is lost by considering the 
isotropic (or transverse 1-dimensional) case $g_{ij}(u)=e(t)^2\delta_{ij}$ 
and correspondingly $A_{ab}(u)=A(u)\delta_{ab}$.\footnote{In particular, such metrics 
can always be elevated to full-fledged 
string theory background by a suitable choice of null dilaton.} 
The Brinkmann and Rosen Lagrangians for a relativistic particle in the lightcone gauge 
$z^+=u=t$ are
\be
\label{ilbcrc}
L_{bc}(z,\dot{z};t) = \trac{1}{2}(\dot{z}^2 -\omega^2(t)z^2)\qquad
L_{rc}(\dot{x};t)=  \trac{1}{2}e(t)^2\dot{x}^2\;\;,
\ee
the general relation 
\eqref{aee} between the Rosen and Brinkmann parameters $g_{ij}$ and $A_{ab}$ 
reducing to the harmonic oscillator equation
\be
\label{Aeo}
A(t)=\frac{\ddot{e}(t)}{e(t)}\equiv-\omega^2(t)\quad \text{or}\quad \ddot{e}(t) + \omega^2(t)e(t)=0\;\;.
\ee
Up to a total derivative term, $L_{bc}$ and $L_{rc}$ are related by the coordinate transformation
$z=e(t)x$,
the former being exactly the Lagrangian of a time-dependent harmonic oscillator, the latter that of a
``free'' particle with time-dependent mass. The non-uniqueness of Rosen coordinates, $x=e(t)^{-1}z$, and of 
the Rosen Lagrangian, is reflected in the
arbitrary choice of solution $e(t)$ of the oscillator equations of motion 
\eqref{Aeo}. 
The corresponding Hamiltonians,
\be
\label{ihbcrc}
H_{bc} = \trac{1}{2}(p_z^2 + \omega(t)^2z^2) \qquad
H_{rc} = \trac{1}{2}e(t)^{-2}p_x^2 \;\;,
\ee
are then related by a linear time-dependent canonical
transformation, the only (barely) non-trivial feature being that
as a consequence of this time-dependence the Hamiltonians differ not
only by the canonical transformation of the canonical variables but
also by the time-derivative of the corresponding generating function
\eqref{gefcantra}.

Brinkmann coordinates provide a global coordinate chart for plane
wave space-times while Rosen coordinates typically (and almost
invariably, e.g.\ for backgrounds satisfying the weak energy condition,
see \cite{gibbonsPW} and Appendix A.1) exhibit spurious coordinate singularities. As a
prototypical example, 
consider the case of constant $A_{ab}$ i.e.\
constant frequency $\omega$. 
In Rosen coordinates one attempts to
use the classical trajectories of the Brinkmann system as Cartesian
coordinate lines (as reflected in the fact that $x^k=\text{const.}$
is always a solution of the Rosen equations of motion), which in the
present case obviously doesn't bode well for these coordinates as the
oscillator reaches its turning point. Indeed, with the choice
$e(t)=\sin\omega t$, say, the Rosen metric and Hamiltonian have the form
\be
ds^2 = 2dudv + \sin^2\omega t\; dx^2\qquad H_{rc} =\trac{1}{2}(\sin^2\omega t)^{-1} p_x^2 
\ee
which is evidently not able to describe the perfectly regular evolution of the system
(or dynamics of strings in e.g.\ the maximally supersymmetric IIB background 
\cite{bfhp1,rrm,mt,bfhp2,bmn}) beyond $t=\pi/\omega$.

The situation may at first appear to be less clear-cut in the case
of genuinely singular plane waves in which both Rosen and Brinkmann
metrics exhibit a singularity at, say, $u=0$. As mentioned in the introduction, 
Rosen metrics are occasionally regarded as null analogues of cosmological
metrics. For instance, isotropic
metrics with $g_{ij}(u)=g(u)\delta_{ij}$ look like null counterparts of
standard spatially flat ($k=0$) FRW metrics, and metrics with a power-law
behaviour $g_{ij}(u)= u^{2a_i}\d_{ij}$ are null analogues of Kasner
metrics. The latter, incidentally, are precisely the Rosen forms of the
scale-invariant plane waves for a certain range of frequencies, 
\be
g_{ij}(u)= u^{2a_i}\d_{ij} \quad\Ra\quad A_{ij}(z^+)= a_i(a_i-1)\d_{ij}\; (z^+)^{-2}\;\;.
\ee
However, interpreting Rosen plane waves in this way requires some care. For
example, consider the simple (both isotropic and power-law) metric with
$g(u)=u^4$. This form of the metric suggests that it describes a null
version of a Big Bang at $u=0$. However, the very same metric can be
written in a different coordinate system as a Rosen plane wave
with $\tilde{g}(u) = u^{-2}$, with the same $u$, 
suggesting that what happens at $u=0$ is a Big Rip, not a Big Bang. 
Clearly then, the connection beween the form of the metric and the 
physics and geometry it actually describes is at best somewhat hidden
in Rosen coordinates. It is however completely manifest in Brinkmann 
coordinates (in which e.g.\ both the above models give rise to a scale-invariant
plane wave with $\omega^2(t) = -2t^{-2}$), since Brinkmann coordinates
are Fermi coordinates along the family of null geodesics with $z^a=0$ and $z^-=\text{const.}$
\cite{fermi}. In particular, the transverse coordinates $z^a$ are Riemann normal
coordinates and thus a direct measure of physical proper distance in the space-time, 
and the wave profile $A_{ab}(z^+) = -R_{+a+b}$ gives the (only non-vanishing) 
components of the Riemann tensor in a parallel propagated frame along this 
null geodesics, and thus directly encodes the only non-trivial geometric 
features of a plane wave in a generally-covariant way.
We will come back to this issue in the context of the corresponding quantum systems in 
section 2.3.

\subsection{Rosen and Brinkmann Quantum States}

For these simple classical systems, the quantum dynamics is evidently
determined completely by the classical dynamics, i.e.\ can be expressed
in terms of solutions to the classical equations of motion. At first
this seems to favour the Heisenberg picture, but it turns out that,
in order to disentangle the time-dependent canonical transformation
from the time-evolution, the Schr\"odinger picture is marginally more
convenient and provides some more insights into the properties of the
Rosen quantum system. Moreover, the proposal of \cite{craps1}, which
is the subject of section 3, is phrased in terms of the (WKB-exact)
WKB wave function of the Brinkmann system. For both these reasons,
we will work in the Schr\"odinger picture.

As generically both the Brinkmann and the Rosen Hamiltonians are time-dependent, 
we will then primarily be interested in the solutions to the time-dependent Schr\"odinger
equations
\be
\label{seq1}
i\del_t \psi_{t}(z) = \hat{H}_{bc} \psi_t(z)\qquad
i\del_t \psi_{t}'(x) = \hat{H}_{rc} \psi_t'(x)
\ee
and in the time-dependence of expectation values of suitably chosen
operators.  Since the canonical transformation between the Brinkmann and
Rosen systems is linear (and we are dealing with a finite-dimensional
quantum system), on general grounds there exists a unitary transformation
\be
U:\mathcal{H}_{rc}\to \mathcal{H}_{bc}:\quad U\psi' = \psi
\ee
from the Rosen to the Brinkmann
space of states implementing this canonical transformation at the quantum
level and mapping solutions of \eqref{seq1} into each other.\footnote{The notation 
throughout will be such that
wave functions in Brinkmann coordinates (Brinkmann states) will be denoted by $\psi(z)$, 
perhaps with other labels indicating individual states or a basis of states, while Rosen
states will be denoted by $\psi'(x)$.
We will not keep track of factors of $2\pi$
in the normalisation of the wave functions.} Its explicit form will not be needed 
in the following but is given in Appendix A.2 for completeness' sake. 

Turning now first to the Rosen system, the Schr\"odinger equation is
\be
\label{rse1}
i\del_t \psi_t'(x) = \hat{H}_{rc}\psi_t'(x) = \frac{1}{2e(t)^2}\hat{p}_x^2
\psi_{t}'(x) 
\ee
In terms of the 
Rosen conformal-time coordinate $T(t)$ (see Appendix A.1),
\eqref{rse1}
reduces to the Schr\"odinger equation of free particle in flat space, 
\be 
\label{rse2}
dT = dt/e(t)^2 \quad\Ra\quad i\del_T \psi_T'(x) = \trac{1}{2}\hat{p}_x^2 \psi_{T}'(x)\;\;.
\ee
Thus a convenient and simple basis of solutions to \eqref{rse1} is provided by the Rosen momentum states
\be
\label{ptkx}
\psi_{t,k}'(x) = \ex{-(ik^2/2)T(t) +i k x}
\ee
(these are evidently eigenstates of $\hat{p}_x$ with eigenvalue $k$, 
delta-function normalised in the usual way).
Note that
the fact that the transformation $t\ra T(t)$ maps the Rosen model evolution to that of a free 
particle in flat space 
already 
shows that in general such a description cannot be valid globally 
(and this reflects the Rosen coordinate singularities).

The time-dependent Schr\"odinger equation of the Brinkmann system 
\be
\label{brsch}
i\del_t \psi_t(z) = \hat{H}_{bc}\psi_t(x) = 
\trac{1}{2}(\hat{p}_z^2 + \omega(t)^2 z^2)\psi_t(z)\;\;,
\ee
on the other hand, 
is simply that of a time-dependent harmonic oscillator. 
Since we have a free (quadratic) theory, the path integral can be explicitly calculated, 
\be
\label{amp}
<z_f|\hat{T}\ex{-(i/\hbar)\int_{t_i}^{t_f}\der t\;\hat{H}(t)}|z_i> =
\frac{1}{\sqrt{2\pi i \hbar(t_f-t_i)}}\sqrt{\frac{\Det [-\del_t^2]}{\Det
[-\del_t^2 -\omega(t)^2]}}\ex{(i/\hbar)S[z_{c}]}\;\;.
\ee
Here $z_c(t)$ now denotes the classical harmonic oscillator solution
with the given boundary condition,
$z_c(t_i)=z_i$ and $z_c(t_f)=z_f$, $S[z_c]$ is the classical action, and
the fluctuation determinants are to be calculated for zero (Dirichlet)
boundary conditions. 

Both the 1-loop
fluctuation determinant and the classical action can be compactly expressed in terms
of the Gelfand-Yaglom (GY) function (see e.g.\ \cite{kleinert} and references therein), 
the solution of the classical equations of motion characterised by
\be
(\del_t^2 + \omega(t)^2)F_{t_i}(t)=0 \qquad 
F_{t_i}(t_i) = 0 \quad \dot{F}_{t_i}(t_i) = 1
\label{gye}
\ee
(and its dual $G_{t_i}(t)=-\del_{t_i}F_{t_i}(t)$, a linearly independent solution
characterised by $G_{t_i}(t_i)=1$, $\dot{G}_{t_i}(t_i)=0$, with Wronskian $W(F,-G)=+1$).
In terms of any linearly independent pair $e_{1,2}(t)$ of the 
classical equations of motion, the GY solution can be constructed as  
\be
\label{gyw}
F_{t_i}(t) = \frac{e_1(t_i)e_2(t) - e_2(t_i)e_1(t)}{W(e_1,e_2)}\;\;.
\ee
For the classical action one then has the expression
\be
\label{sxc}
S[z_c]= \frac{1}{2F_{t_i}(t)}[\dot{F}_{t_i}(t) z^2 +
G_{t_i}(t) z_i^2 - 2 z z_i]\;\;,
\ee
and for the ratio of fluctuation determinants one has the remarkably simple result
\be
\frac{1}{\sqrt{2\pi i \hbar(t_f-t_i)}}\sqrt{\frac{\Det [-\del_t^2]}{\Det
[-\del_t^2 -\omega(t)^2]}} =
\frac{1}{\sqrt{2\pi i \hbar F_{t_i}(t_f)}}\;\;.
\label{gyratio}
\ee
The above-mentioned properties
of the GY function and its dual imply that it also governs
the time-evolution of the classical system, since one has
\be
\label{gyu}
\begin{pmatrix}
z(t)\\
\dot{z}(t)
\end{pmatrix}
=
\begin{pmatrix}
G_{t_i}(t) & F_{t_i}(t) \\
\dot{G}_{t_i}(t) & \dot{F}_{t_i}(t)
\end{pmatrix}
\begin{pmatrix}
z(t_i)\\
\dot{z}(t_i)
\end{pmatrix}
\ee
The amplitude \eqref{amp} therefore gives rise to the family of Brinkmann propagator (or WKB) states
\be
\label{ptzz}
\psi_{t,z_i}(z) \sim \frac{1}{\sqrt{F_{t_i}(t)}}\ex{i S[z_c(z,z_i;t,t_i)]}
\ee
characterised by $\lim_{t\ra t_i}\psi_{t,z_i}(z) = \delta(z-z_i)$.
Thus the GY function compactly encodes much of the relevant information about the 
quantum system, and for this reason we will focus on the behaviour of the GY function
(and expectation values of certain operators) in our analysis of singular systems in
section 3.

\subsection{Brinkmann vs Rosen expectation values}

Turning now to expectation values of quantum operators, a particularly simple (and easy to work with)
basis of states is provided by the Rosen momentum states $\psi'_{t,k}$ \eqref{ptkx}. Thus we consider
general wave packets
\be
\label{rwp}
\psi_t'(x) = \int \der k\;a(k)\psi'_{t,k}(x)
\ee
(with the assumption that the $a(k)$ are square integrable and differentiable)
and their Brinkmann duals $\psi = U\psi'$. For position observables, Rosen and Brinkmann
expectation values are connected in a simple way, $U^{-1}\hat{z}U = e(t) \hat{x}$
\eqref{qct} implying relations like
\be
\label{z2}
<\psi|\hat{z}^2|\psi> = e(t)^2 <\psi'|\hat{x}^2|\psi'>\;\;.
\ee
Moreover, since the Rosen momentum states are eigenstates of the momentum operator $\hat{p}_x$, 
it is straightforward to determine the expectation value of the Rosen Hamiltonian
in such a wave packet, 
\be
\label{hrcvev}
<\psi'|\hat{H}_{rc}|\psi'> = \frac{1}{2e(t)^2}\int \der k\; k^2 a(k)^* a(k)\;\;.
\ee
We note that if the wave packet $\psi'_{t,x}$ is to be a solution of the Schr\"odinger equation, 
the coefficients $a(k)$ need to be time-independent. Thus we conclude that the time-dependence
of the expectation value of the Rosen Hamiltonian is determined entirely by the prefactor
$e(t)^{-2}$ of the Hamiltonian. 
In particular, if $e(t)$ has a zero somewhere, $e(t_0)=0$, the
expectation value of the Rosen Hamiltonian diverges as $t\ra t_0$
for any solution to the Schr\"odinger equation. We will come back to this and related issues
below.   

In order to calculate (and compare this with) the expectation value of $\hat{H}_{bc}$, one needs
to take into account the non-trivial relationship between the classical Hamiltonians provided
by the generating function \eqref{gefcantra}. Relegating the details of the calculations to the
appendix, the results for the expectation values of $\hat{H}_{rc}$, $\hat{x}^2$ and their 
Brinkmann counterparts can be summarised as
\be
\label{abc1}
\begin{aligned}
\langle \psi'|\hat{x}^2|\psi'\rangle&= \mathbb{A}+ T(t) \mathbb{C} + T(t)^2 \mathbb{B}\\
\langle\psi |\hat{z}^2 |\psi\rangle&= e(t)^2 \mathbb{A} + \tilde{e}(t)^2 \mathbb{B} +
e(t)\tilde{e}(t)\mathbb{C}\\
\langle\psi'|\hat{H}_{rc}|\psi'\rangle&= \trac{1}{2}e(t)^{-2}\mathbb{B}\\
\langle\psi|\hat{H}_{bc}|\psi\rangle&= E(e)(t)\mathbb{A} + E(\tilde{e})(t)\mathbb{B} +
E(e,\tilde{e})(t)\mathbb{C} 
\end{aligned}
\ee
where the (constant) coefficients $\mathbb{A}\neq 0,\mathbb{B}\neq 0,\mathbb{C}$, defined
in \eqref{ABC}, depend only on the modes $a(k)$ of the wave packet, and the functions 
\be
E(e)=  \frac{1}{2}(\dot{e}^2+\omega^2e^2)\quad\text{and}\quad E(\tilde{e})=
\frac{1}{2}(\dot{\tilde{e}}^2+\omega^2\tilde{e}^2)
\ee
can be readily intepreted as the
classical energies of the solutions $e(t)$ and $\tilde{e}(t)$. We have also introduced
$E(e, \tilde{e})= 1/2(\dot{e}\dot{\tilde{e}}+\omega^2e\tilde{e})$ for uniformity of notation, 
and we have used some identities satisfied by $e(t)$ and its dual linearly independent
solution $\tilde{e}(t)=e(t)T(t)$ to put the results for the Brinkmann expectation values
into a form which makes it manifest that it
is symmetric under the exchange $e(t)\lra\tilde{e}(t)$.\footnote{Recall that, unlike the Rosen system, 
the Brinkmann system depends only on the frerquency $\omega(t)$, not on a particular
classical solution $e(t)$, and this should be (and is of course) reflected in the expectation values etc.}

None of the results \eqref{abc1} should come as the slightest
surprise since they just mirror the time-dependence
of the corresponding general classical solution, a linear combination of $e(t)$ and its
dual $\tilde{e}(t) = e(t)T(t)$.
In particular, this gives the expected result for the expectation value of the Hamiltonian of the 
usual time-independent harmonic oscillator with constant frequency $\omega$. In this
case one has e.g.\ 
$e(t)=\sin\omega t$ and $\tilde{e}(t)= -\omega^{-1}\cos\omega t$ so that
\be
\begin{aligned}
e(t) = \sin\omega t &\quad\Ra\quad E(e)=\trac{1}{2}\omega^2\quad,\quad E(\tilde{e}) = \trac{1}{2}
\quad,\quad E(e,\tilde{e})=0\\
&\quad\Ra\quad
\langle\psi|\hat{H}_{bc}|\psi\rangle =\trac{1}{2}\int \der\ell\; \left[ \omega^2 |a'(\ell)|^2 + \ell^2 |a(\ell)|^2\right]
\end{aligned}
\ee
For the ground state wave function, with $a(\ell)\sim \exp(-\ell^2/2\omega)$, one finds
$\langle\psi|\hat{H}_{bc}|\psi\rangle= \omega/2$. 

Note that the corresponding Rosen Hamiltonian, on the other hand, 
fails completely to exhibit this perfectly regular behaviour of the time-independent harmonic
oscillator. As noted before, the expectation value of the Rosen Hamiltonian diverges for all states
as $t\ra t_0$ with $e(t_0)=0$, and for any solution of the harmonic oscillator equations 
there is an infinite number of such $t_0$.

Geometrically, this is a pure coordinate singularity of the Rosen metric,
and the quantum system fails to detect that this is not an honest
singularity.  The problem is not so much the Rosen states but the operators
whose expectation values one wants to evaluate. To highlight this,
consider the statements
\be
<\psi|\hat{H}_{bc}|\psi> = <\psi'|(\hat{H}_{bc})_{rc}|\psi'>
\qquad
<\psi'|\hat{H}_{rc}|\psi'> = <\psi|(\hat{H}_{rc})_{bc}|\psi>
\ee
that follow from \eqref{obc}.\footnote{$(\hat{H}_{bc})_{rc}$ denotes the 
Brinkmann Hamiltonian expressed in terms of the
Rosen operators $\hat{x}, \hat{p}_x$ etc. Note that this is not the same as the Rosen
Hamiltonian $\hat{H}_{rc}$, cf.\  \eqref{qhh} for the relation between the two.}
In the time-independent case, say, of a harmonic oscillator with
constant real frequencies, the Brinkmann expectation value of the
Brinkmann Hamiltonian is of course well-defined and non-singular
in every Schr\"odinger state.  Therefore, by the above identity,
also the expectation value of the corresponding Rosen operator
$(\hat{H}_{bc})_{rc}$ is non-singular and well-defined in every one
of the corresponding Rosen states - it is a good Rosen operator. The
Rosen Hamiltonian $\hat{H}_{rc}$, on the other hand, is not: it has a
divergent expectation value in any Rosen state, and its Brinkmann-picture
counterpart $(\hat{H}_{rc})_{bc}$ correspondingly has a divergent
expectation value in every Brinkmann state.

The fact that the Brinkmann versus Rosen issue is a question of operators
and not of states is also reinforced by the space-time origin and
interpretation of these models, which provide additional physical criteria
allowing one to select the operators one attributes physical significance
to, thus going beyond the mere quantum mechanical formalism.

As an example, let us consider the Rosen expectation value
$<\hat{x}^2>$. This gives the mean value of the Rosen coordinate-distance
squared in some state. Without any further physical input, one can argue
until one is blue in the face whether this is the physically relevant
quantity to look at or whether one should rather look at $<\hat{z}^2>$.
Indeed, in principle nothing stops one from
constructing a non-relativistic quantum system with a Rosen Hamiltonian,
and with the Rosen coordinates $x^i$ Cartesian flat space coordinates,
and in this case indeed it would be $<\hat{x}^2>$ that is the relevant quantity 
that can be related to observations and measurements of the system.

However, recalling the origin of the Rosen and Brinkmann Hamiltonians as the 
lightcone Hamiltonians of relativistic particles in the plane wave background,
albeit in different coordinate systems, the situation presents itself in a
rather different light. Indeed, in this case 
physical systems detect and measure not
coordinate distance but generally-covariant
quantities like proper distance, and this can easily be accounted for, 
\be
<(\text{proper Rosen distance})^2> = <e(t)^2\hat{x}^2>\;\;.
\ee
But according to \eqref{z2}, this is just the Brinkmann expectation value of $\hat{z}^2$, 
\be
<\psi'|(\text{proper Rosen distance})^2|\psi'> = <\psi|\hat{z}^2|\psi>\;\;.
\ee
In this case, even if one starts in the Rosen system one is invariably
led to the Brinkmann expectation value as the physically well-defined
quantity to look at.

Since the time-evolution of this physically meaningful notion of distance
is governed by the Brinkmann Hamiltonian and not by the Rosen Hamiltonian
($d\hat{z}/dt \sim [\hat{H}_{bc},\hat{z}]$),
this shows that it is also the Brinkmann and not the Rosen Hamiltonian that
is a physically meaningful operator in the quantum theory of a relativistic 
particle.

We will see below that this conclusion can be singificantly strengthened
in that the Brinkmann dynamics is also singled out in the dynamics of
strings and fields by the requirement of unitarity (and that the Rosen
dynamics cannot be unitarily implemented on any field theory Fock space).

\subsection{Unitarity and Uniqueness at the QFT/String Level\\
(Brinkmann wins hands down)}

So far we have studied aspects of the quantum mechanics of point
particles in plane wave backgrounds. When one considers perturbative
string theory, the relevant dynamics of closed strings (in the lightcone gauge) is that
of the transverse string modes on the cylinder $S^1 \times \mathbb{R}$, and it is well
known that in Brinkmann variables this reduces to the standard free action
of massive scalar fields with time-dependent mass matrix $-A_{ab}(t)$
(the frequency/profile of the Brinkmann plane wave). The same is true
for non-perturbative matrix string theory: in \cite{mols}
it was shown that the near-singularity $t\to0$ evolution of the plane
wave matrix big bang models of \cite{mmpwbb} is dominated by the
same time-dependent mass terms, while the quartic interaction and the
accompanying non-Abelian nature of the dynamics are subleading.
In the isotropic case we have focussed on one is thus in both cases confronted with the
problem of the quantisation of a single scalar field on $S^1 \times \mathbb{R}$
with time-dependent mass term $m^2(t) = \omega^2(t)$ (or with an 
infinite number of Fourier modes with masses $m_n(t)^2 = m(t)^2 + n^2,\; n \in\ZZ$).

It is well known that in the quantisation of systems with an infinite
number of degrees of freedom (for which is there is no analogue
of the Stone - von Neumann theorem) one is confronted with various
ambiguities, e.g.\ in the choice of field parametrisation or, for a fixed
parametrisation, in the choice of (Fock) representation of the canonical
commutation relations (CCRs). In the usual setting of a Poincar\'e-invariant
field theory (of a free scalar field, say), the requirement of Poincar\'e
invariance of the Fock vacuum selects a unique Fock representation of the
CCRs on which the dynamics can be unitarily
implemented. In general (in particular non-stationary) space-times there
is no such unique choice, and this give rise to phenomena like particle
production etc. 

When there are some spatial isometries, it is natural to require that
the quantisation procedure respect these symmetries, but (a) even in
spatially maximally symmetric (cosmological) situations this is not enough
to uniquely fix the representation, and (b) this says nothing about the
dynamics and the possibility to unitarily implement the time-evolution.

In a series of papers, motivated by cosmological considerations, Cortez
et al.\ (see e.g.\ \cite{cortez1,cortez2,cortez3,cortez4,cortez5})
investigated whether and to which extent the joint conditions
of (a) invariance under the background symmetries and (b) unitary
implementation of the dynamics can be used to select and specify a unique
Fock quantisation. In particular, they
investigated the case of scalar fields on $S^d \times \mathbb{R}$, $d\leq 3$,
with a
time-dependent mass (precisely the situation we are interested in here
for $d=1$), with some rather remarkable and striking uniqueness results.

To describe these, consider a scalar field $\phi$ satisfying the equation
\be
\label{ceom}
\ddot{\phi}-\Delta\phi + s(t)\phi = 0
\ee
where $\Delta$ is the Laplacian on $S^d$ and $s(t)$ is the mass-squared function
satisfying some mild regularity condition ($C^1$ on the time interval $I$
under consideration being sufficient). In particular, no condition on the 
sign of $s(t)$ is (or needs to be) imposed.\footnote{A caveat in e.g.\ \cite{cortez3,cortez5}
regarding such
a condition for the quantisation of the zero mode sector is unnecessary; the 
quantum theory of a particle with potential $s(t)z^2$, say, is well-defined
for any sufficiently regular $s(t)$, regardless of its sign; one should simply
not try to define it via naive analytic continuation from the usual Schr\"odinger
theory of a harmonic oscillator with $\omega^2 > 0$ as that would lead to a 
non-hermitian Hamiltonian.}

$SO(d+1)$-invariance is now imposed
as a condition on the complex structure defining the creation and annihilation
operators $(\hat{a}_n,\hat{a}_n^*)$, thus on the Fock representation of the CCRs (this 
still leaves a lot of possibilities). The requirement of unitary 
implementability of the dynamics is then tantamount to the condition
that the Bogoliubov transformation 
\begin{equation}
\label{bog1}
\left(\begin{array}{c} \hat{a}_n(t)\\ 
\hat{a}^*_n(t)\end{array}\right)=\left(\begin{array}{cc} \alpha_n(t,t_i) & \beta_n(t,t_i)
\\ \beta^*_n(t,t_i) &  
\alpha^*_n(t,t_i) \end{array}\right)\left(\begin{array}{c} \hat{a}_n(t_i)\\ \hat{a}^*_n(t_i)\end{array}\right)
\end{equation}
describing the temporal evolution $t_i\ra t$ of some suitably defined instantaneous creation and annihilation
operators leads to finite ``particle production'', $\sum_n|\beta_n|^2 < 0$.
Then the first main result is the following:
\begin{enumerate}
\item For every $s(t)$ (subject to the above-mentioned regularity assumptions), up to unitary 
equivalence there exists a unique $SO(d+1)$-invariant 
Fock representation of the CCRs (complex structure) with respect to which the dynamics is unitarily
implemented. This representation is independent of $s(t)$ and, in particular, coincides with the 
standard Fock representation of a free massless field in Minkowski space. 
\end{enumerate}
Note that spatial compactness is crucial for the validity of the 2nd statement, IR problems 
preventing an analogous statement from being true e.g.\ for a free scalar field in Minkowski space, where the 
Fock represenations (and corresponding representations of the Poincar\'e group) are
of course not unitarily equivalent for different masses.

In the above, we started with a fixed parametrisation of the fields, the scalar field $\phi$ and its
canonically conjugate $P_{\phi}$ (a spatial density on $S^d$). The second main result of Cortez et al.\
\cite{cortez5} 
concerns the ambiguity resulting from time-dependent linear canonical transformations (field redefinitions) 
of the form
\be
\label{ctf}
\varphi=F(t)\phi \quad , \quad 
P_\varphi=P_\phi/F(t)+G(t)\sqrt{h}\phi
\ee
where $h$ is the determinant of the spatial metric, 
$F(t)$ is different from zero everywhere and both $F(t)$ and $G(t)$
are differentiable. Such transformations are commonly performed e.g.\ in the 
context of cosmological perturbation theory where they are used to put the
action for the perturbations into a suitable (canonical) form.
This transformation has 
an obvious and simple effect on the above Bogoliubov transformation implementing
the time evolution, and one can analyse the conditions arising from the 
square-summability of the new $|\beta_n|$. It turns out that the joint requirements
of symmetry and unitarity also determine an essentially unique field parametrisation:
\begin{enumerate}
\addtocounter{enumi}{1}
\item The dynamics cannot be unitarily implemented in an $SO(d+1)$-invariant Fock representation
unless $F(t)$ and (for $d\neq 1$) $G(t)$ are constant.
\end{enumerate}
When $d=1$, shifts $P_\phi\ra P_\phi + G(t)\phi$ of the momentum are allowed, all leading to the
same unique (free massless) representation of the CCRs singled out in the first result. 
We will see below that, for the point transformation that arises when going
from Brinkmann to Rosen coordinates, $F(t)$ constant already all by itself 
implies $G(t)=0$ so that this is, in any case, not an issue.

Intuitively, the fact that there is a unique field parametrisaton that allows a unitary 
implementation of the dynamics can be attributed to the fact that any non-trivial
time-dependent reparametrisation of the field will give rise to damping/friction terms
in the equations of motion which are obstructions to unitarity \cite{cortez5}.

These results have several implications for the study of (matrix) string
theory in plane wave backgrounds, in particular as regards the issue of
whether to use a Brinkmann or a Rosen parametrisation of the fields, 
in which the scalar sector of the action schematically takes the form
\be
S_{bc}=-\trac{1}{2}\int \der^2\sigma 
\left(\eta^{\alpha\beta}\del_\alpha Z^a\del_\beta Z^b -A_{ab}(t)Z^aZ^b + \ldots\right)
\ee
or
\be
S_{rc}=-\trac{1}{2}\int \der^2\sigma 
\left(\eta^{\alpha\beta}g_{ij}(t)\del_\alpha X^i\del_\beta X^j + \ldots\right)\;\;,
\ee
the $(+\ldots)$-term including possible interaction terms and/or couplings to background fields.
Exactly as for the point particle action (Appendix A.1), up to a total derivative these two actions 
are related by the time-dependent scaling $X^i = e^i_a(t)Z^a$ with $g_{ij}(t)e^i_a(t)e^j_b(t)=\d_{ab}$
(this equivalence actually extends to the full non-Abelian matrix string or 3-algebra multiple
M2-brane action \cite{mmM2}).
In the isotropic case $g_{ij}(t)=e(t)^2 \d_{ij}$, this corresponds to the canonical
transformation \eqref{ct}
\be
\label{ctzx}
X = Z/e(t)\quad,\quad P_X = e(t)P_Z - \dot{e}(t)Z\;\;,
\ee
which is precisely of the form \eqref{ctf} with $F(t)=1/e(t)$ and $G(t)=-\dot{e}(t)$
(in particular, in this case one has $F(t)=\text{const.} \Ra G(t)=0$).
Note that the Brinkmann field $Z$ has a canonical kinetic term and therefore satisfies the 
canonical equations of motion \eqref{ceom}, while the Rosen field $X$ plays the role of
the reparametrised field $\varphi$. 

The implications of the results of Cortez et al.\ are now clear. We
consider frequencies / mass terms $s(t)=\omega(t)^2=-A(t)$ and a time
interval $I$ such that the assumptions about $s(t)$ are satisfied (and
will comment on the implications for more singular situations in section 3).
\begin{enumerate}
\item In the Brinkmann parametrisation there is a unique Fock quantisation 
respecting the translation invariance along the worldsheet $S^1$
(namely the standard Fock space of a free massless scalar field) 
such that the dynamics is unitarily implemented.
\item 
There is no $S^1$-invariant Fock quantisation of the Rosen system 
with unitary time evolution.
\end{enumerate}
A priori these results do not exclude the possibility that, starting 
from scratch in Rosen coordinates, one could find an exotic Fock 
representation of the CCRs and implement unitarily the temporal evolution. We 
know for sure, though, that the vacuum of such a representation would 
not be invariant under $S^1$ rotations. The other alternative would be a non-Fock
representation. Neither option seems palatable. We believe that this should 
settle once and for all the question whether one should use Brinkman or Rosen fields,
and we will focus entirely on the Brinkmann systems in the following.

\section{Regularisation of the Dynamics for Singular Plane Waves}

We will now consider the quantum models arising from Brinkmann plane
wave metrics \eqref{mpwrc} with profile $A_{ab}(z^+)\sim (z^+)^{-2}$,
as these were shown in \cite{bbop1,bbop2,gpsi} to arise generically
as the Penrose limits of metrics with singularities, and to which
the derivation of \cite{csv} of a matrix string model (the matrix big bang)
was extended in \cite{mmpwbb}. 
The frequencies that typically
arise from the Penrose limit procedure 
have the form $a(1-a)(z^+)^{-2}$
for some $a \in \mathbb{R}$ and continuing to focus on the isotropic
case we will thus specifically study the models with frequency / 
mass terms
\begin{equation}
\label{masst}
\omega^2(t) =\frac{a(1-a)}{t^2}\;\;.
\end{equation}
The problem is evidently invariant under $a \ra 1-a$ and we can therefore, without loss of generality, 
restrict to the range $a> 1/2$.\footnote{The value $a=1/2$ would require a separate treatment, because one
of the classical solutions has a logarithmic $\sim t^{1/2}\log t$ rather than a power-law behaviour, 
but is not particuarly interesting in other respects. Aspects of the $a=1/2$ model (referred to 
as the Gowdy model in the cosmological context)
are analysed in detail in \cite{cortez1}.} We will mainly be interested in the range $a>1$
for which $\omega^2(t)$ is negative, since it is this tachyonic behaviour that characterises
the Penrose limits of string theory backgrounds with strong string coupling singularities \cite{mmpwbb}
(for which a non-perturbative description such as matrix string theory thus also becomes mandatory),  
but many of our considerations are also valid in the weakly coupled range $1/2 < a < 1$
which was the focus of interest e.g.\ in \cite{prt}.

We recall and reiterate here that it was argued in \cite{mols}, on the
basis of numerical and analytical investigations of various toy-models,
that both the classical and the quantum evolution of the matrix (string)
system near strong string coupling (and hence weak gauge coupling)
singularities are generically driven entirely by the divergent tachyonic mass
terms. In particular, the quartic interaction potential, which comes
with a time-dependent coefficient $g_{YM}^2 \sim t^{2q}$ with $q>0$, can
be shown to be self-consistently small in quantum-mechanical perturbation
theory in that regime and appears to (perhaps somewhat disappointingly)
play no role there. Thus, in the following we can and will focus on
the linear harmonic oscillator dynamics even when what we have in mind are 
primarily applications to matrix string theory.

\subsection{Necessity of Regularisation}

The considerations of section 2.4 imply that there will be a unique
Brinkmann quantisation of the system with a unitary time evolution as
long as one stays away from $t=0$ (since \eqref{masst} satisfies the
required regularity assumptions for $t \geq t_0 > 0$).

On the other hand, it is self-evident that the quantum theory with
frequency / mass terms will not miraculously regularise the classical
singularity of the theory as $t\ra 0$, since the evolution of wave packets
and expectation values of this WKB-exact quantum system will track that
of the singular classical system. Nevertheless, we will now explicitly
display some of the resulting singularities, for reference purposes,
and since we will attempt to regularise these expressions below.

Starting from the fact that the two linearly independent
solutions of the classical equations of motion 
\be
\ddot{e}(t) + a(1-a) t^{-2} e(t) = 0
\ee
of the point particle system, defined for $t>0$, can be chosen to be
\be
e(t) = t^{1-a}\quad,\quad \tilde{e}(t) = t^a/(1-2a)
\ee
(with unit Wronskian),
one finds from \eqref{gyw} that the Gelfand-Yaglom function is 
\be
\label{gyshpw}
F_{t_i}(t) = \frac{1}{1-2a} (t_i^a t^{1-a} - t_i^{1-a}t^a)\;\;.
\ee
A priori this is not defined for $t_i$ or $t$ negative. Moreover,  
this expression evidently diverges for all $t>0$ as $t_i\ra 0$ for $a>1$
and goes to zero for all $t>0$ as $t_i\ra 0$ for $1/2<a<1$, either behaviour
indicating a breakdown of the quantum theory. 
Note that the appearance of an isolated zero of the GY function for some 
$t=t_f$, $F_{t_i}(t_f)=0$, is not a fundamental problem.
In view of \eqref{gye},  this just signals the existence of a
zero mode of the Dirichlet problem which needs to (and can) be taken
care of in any number of standard ways. What happens here is that
the GY function goes to zero (or diverges) for all $t>0$, and this
is a genuinely singular behaviour.

One can also see that unitarity of the time-evolution breaks down as
$t\ra 0$, signalled by the divergence of the Bogoliubov coefficient
$\beta_{n=0}(t,t_i)$ in \eqref{bog1} as $t_i\ra 0$, leading a fortiori to a breakdown
of unitarity of the full quantum field / string theory. This can be
established either by transforming the GY time-evolution matrix \eqref{gyu}
to an appropriate oscillator basis, or directly by a calculation of the Bogoliubov
coefficient $\beta(t,t_i)$ between instantaneous eigenstates at times $t$ and $t_i$.\footnote{For 
$a \neq 0,1$, $\beta(t,t_i)$ turns out to have the general form 
``$(\text{terms} \sim t^{-a}, t^{1-a}) \times (\text{terms}\sim
t_i^{a-1}, t_i^{a})- (t\lra t_i)$''. 
Thus as $t_i\ra 0$, $\beta(t_1,t_2)$ diverges for any $a\neq 0,1$ (for $a=0,1$ one just has 
a free particle).}


Finally, for Brinkmann expectation values one finds from \eqref{abc1} that
schematically
\be
\label{abc2}
\begin{aligned}
\langle \psi|\hat{z}^2|\psi\rangle&\sim \mathbb{A} t^{2-2a}+\mathbb{B} t^{2a} + \mathbb{C} t\\
\langle \psi|\hat{H}_{bc}|\psi\rangle &\sim \mathbb{A} t^{-2a}+\mathbb{B} t^{2a-2} + \mathbb{C} t^{-1}\;\;,
\end{aligned}
\ee
which are also all singular as $t\ra 0$ (suppressed numerical factors hidden in the $\sim$ signs making
the coefficients of the potentially divergent terms zero only for $a=0,1$, i.e.\ $\omega^2(t)=0$). 

\subsection{Single Scale Regularisation, Gelfand-Yaglom and Fine-Tuning}

We will now try to make (at least some of) the above expressions well defined by 
a regularisation of the frequency / plane wave profile
(thus maintaining the ($\alpha'$-exact) plane wave character of the background metric).
Adopting at first the suggestion in \cite{craps2} we will explore the possibility of using a 
single-scale regularisation of the form
\be
\label{ssr}
\omega^2(t) \ra 
\Omega_{\epsilon}^2(t) = \epsilon^{-2}\Omega^2(t/\epsilon) \equiv  \epsilon^{-2}\Omega^2(\eta) 
\ee
in which the scale invariance of the original background is as much as possible
respected and the classical equation of motion becomes independent of $\epsilon$, 
\be
\label{Feq}
\frac{d^2}{d\eta^2} f(\eta) + \Omega^2(\eta)f(\eta) =0 \;\;.
\ee
In order to qualify as a regularisation, $\Omega_\epsilon^2(t)$ should be non-singular for all $t$, 
in particular also $t=0$, as long
as $\epsilon\neq 0$, and should reduce to $\omega^2(t)$ as $\epsilon\ra 0$ for $t> 0$. In terms of
$\Omega^2(\eta)$ this amounts to the conditions
\be
\exists \; \lim_{\eta\ra 0}\Omega^2(\eta) \equiv \Omega^2(0)\quad\text{and}\quad
\Omega^2(\eta) \ra a(1-a) \eta^{-2} \quad\text{for}\quad \eta \ra \infty
\ee
We will moreover assume that 
for $\epsilon\ra 0$ and $t\neq 0$ one reaches the same limiting frequency $a(1-a)t^{-2}$
for $t<0$ as for $t>0$ (the case of different left/right frequencies having already been
dismissed in \cite{craps2}).
In that case we also have the 
same asymptotic behaviour of $\Omega^2(\eta)$ as $\eta\ra-\infty$.

Some examples of regularisations satisfying these conditions that come to mind immediately 
are 
\be
\label{reg1}
\Omega_\epsilon^2(t) = a(1-a) (t^2 + \epsilon^2)^{-1} \quad\LRa\quad \Omega^2(\eta) = a(1-a)(\eta^2+1)^{-1}
\ee
and its generalisations
\be
\label{reg2}
\Omega^2(\eta) = a(1-a)\frac{\eta^{2n}}{(\eta^2+1)^{n+1}}
\ee
(all of which we will unfortunately have to dismiss below for not accomplishing what they are supposed to do).
Note that these examples have the property that $\Omega^2(-\eta)=\Omega^2(\eta)$, so that
the regularised frequency preserves the $\mathbb{Z}_2$
reflection symmetry $t\ra -t$ of the unregularised frequency. This is a natural additional assumption, but one
which will not play a significant role in the analysis below.

We can then assume (see also \cite{oetn}) that we have two linearly independent and $\epsilon$-independent
solutions $f_k(\eta)$ of \eqref{Feq} with the asymptotic behaviours
\be
\label{Fas}
\bpm f_1 \\ f_2 \epm \stackrel{\eta\ra\pm\infty}{\longrightarrow} \bpm \alpha_\pm & \beta_\pm \\
\gamma_\pm & \delta_\pm \epm \bpm |\eta|^{a} \\ \;\;|\eta|^{1-a} \epm
\ee
(Wronskian conservation implying that the ratio of the determinants of the two matrices is $-1$).
Note that the generic behaviour of a solution as $|\eta|\ra\infty$ is the dominant
behaviour $f\sim |\eta|^{a}$ (recall that we have chosen to consider the range $a> 1/2$).
We will call a solution 
\textit{fine-tuned} if it approaches the subleading solution $|\eta|^{1-a}$
for both $\eta \ra +\infty$ and $\eta \ra - \infty$. The existence of such a solution is not 
guaranteed.  If it exists, it has the following properties:
\begin{itemize}
\item If a fine-tuned solution exists, it is unique (any other 
linearly-independent solution has to exhibit the dominant asymptotic
behaviour $\sim |\eta|^a$ for a non-zero Wronskian at $\pm\infty$).
\item
If the regularised potential is even, then the fine-tuned solution is either even or odd
(since otherwise one could construct another fine-tuned solution by taking its even or odd part).
\end{itemize}
From this pair of linearly independent solutions we can constuct the Gelfand-Yaglom function
\be
\label{gyeps}
F^\epsilon_{t_i}(t) =
W(f_1,f_2)^{-1}\left(f_1(t_i/\epsilon)f_2(t/\epsilon)-f_2(t_i/\epsilon)f_1(t/\epsilon)\right)
\ee
and we want to investigate its behaviour as $\epsilon\ra 0$. By construction this limit is well defined
if $t_i,t$ are both positive, giving rise to \eqref{gyshpw}.
The main result is that the limit is also well defined for $t_i<0, t>0$ iff a fine-tuned solution exists, 
\be
\label{Fft}
t_i<0, t>0: \quad \exists \lim_{\epsilon\ra 0} F^\epsilon_{t_i}(t)\quad\LRa\quad \exists \quad \text{fine-tuned solution}
\ee 
Proof: in evaluating this limit using the asymptotic behaviour, and keeping in mind that 
a constant $\eta$-Wronskian $W_\eta(f_1,f_2)$ implies that the relevant $t$-Wronskian,  
$t=\epsilon\eta$, is 
$\sim \epsilon^{-1}$, one finds one potentially divergent 
contribution $\sim \epsilon^{1-2a}$ whose coefficient is $\alpha_+\gamma_--\alpha_-\gamma_+$. But
\be
\alpha_+\gamma_--\alpha_-\gamma_+=0 \quad\LRa\quad \exists A,C: 
\left\{\begin{array}{c} A\alpha_+ + C\gamma_+=0\\
A\alpha_- + C\gamma_-=0 \end{array}\right. \quad\LRa\quad Af_1 + Cf_2 \;\text{fine-tuned}
\ee
This is a reformulation of a statement in \cite{oetn,craps2} (where the GY function 
is called the compression factor).
With this condition in place, one finds that the regularised GY function is finite as $\epsilon\ra 0$.
Assuming without loss of generality that in \eqref{Fas} it is $f_2$ that 
is the fine-tuned solution (so that $\gamma_\pm=0$ and $\alpha_+ \alpha_-\neq 0$), one finds
\be
\label{GYcontinuations}
\lim_{\epsilon\ra 0} F_{t_i}(t) = \frac{1}{1-2a}
(q |t|^a |t_i|^{1-a} + q^{-1} |t_i|^a |t|^{1-a})\;\;,
\ee
where $q=\alpha_+/\alpha_-$. Whenever the fine-tuned solution $f_2$ can be chosen to
be of definite parity (e.g.\ when the regularised profile is even), one has $|q|=1$
($q=\pm 1$ for $f_2$ odd/even). 
Note that this is quite different from a naive analytic continuation 
of \eqref{gyshpw} through $t=0$, which (for non-integer $a$) 
would give a relative imaginary phase between the two terms, and from
the result one obtains \cite{prt}
from the analytic continuation of the Bessel functions appearing as
solutions to the string mode equations in this background.

This raises the questions how restrictive the condition is that the
regularised frequency admits a fine-tuned solution and how to characterise
the regularised profiles that admit such a fine-tuned solution. We will
now provide some partial answers to these questions.

We consider the range $a>1$ for which $\omega^2(t)=a(1-a)t^{-2} <0$.
First of all, we observe that a fine-tuned solution, behaving
as $|\eta|^{1-a}$ for $|\eta|\ra\infty$, initially (starting from
$\eta=-\infty$) grows, accelerating upwards, the unregularised solution
diverging as $\eta\ra 0_-$. In the regularised case, however, in order to
regain the fine-tuned behaviour as $\eta\ra +\infty$, the solution has to
turn around all the way and bend down (decelerate). This requires that
there is a region around $\eta =0$ in which the regularised profile
$ \Omega_{\epsilon}^2(t)$ changes sign and becomes positive. A simple
(and rather crude) regularisation accomplishing this can be obtained 
by patching the original profile outside a given $\eta$-interval to a constant potential for
$\eta\in [-\eta_0,+\eta_0]$. In terms of the step-function $H(\eta)$
this means
\be 
\label{toy1}
\Omega^2(\eta) = a(1-a) \eta^{-2} H(|\eta|-\eta_0) + \lambda^2 H(\eta_0 - |\eta|) 
\ee
with $\lambda^2$ a dimensionless constant.
In terms of the original variable $t$ the second term corresponds to a potential wall 
of width $2\epsilon\eta_0$ and height $\lambda^2/\epsilon^2$. A standard exercise
in matching conditions reveals that for any value of 
the parameter $a$ this potential admits a fine-tuned solution 
for an appropriate choice of the parameter $\lambda$. 

On the other hand, the above positivity criterion immediately rules
out all the candidates \eqref{reg1} and \eqref{reg2} since for $a>1$
they are strictly negative for all $\eta$.  While there appears to
be no analogous obvious (positivity or other) criterion for the range
$0<a<1$, the range of principal interest in \cite{craps2} (and
also previously e.g.\ in \cite{prt}), it is nevertheless true that
\eqref{reg1} possesses no fine-tuned solutions in that range either,
as can be seen by explicitly solving the equations of motion in terms
of hypergeometric functions (cf.\ also the comment in footnote 36 of
\cite{prt}). Thus for this class of examples the claim in \cite{oetn}
(translated into our terminology) that generically fine-tuned solutions
should exist for a discrete spectrum of values of $a$ is confirmed in
the rather un-useful way that this spectrum is empty.

This shows that it is not completely obvious how to find or determine regularisations that
admit the required fine-tuned solutions. In order to side-step this issue, we turn 
the question
on its head by starting with a fine-tuned regularisation $e^\epsilon(t)$ of 
the solution $e(t)=t^{1-a}$ (and thus of the Rosen metric) 
and determining the corresponding regularised 
frequency $\Omega_\epsilon^2(t)$,  
\be
e(t)=t^{1-a}\ra e^{\epsilon}(t) \quad\Ra \quad \Omega_\epsilon^2(t) =
-\frac{\ddot{e}^\epsilon(t)}{e^\epsilon(t)}\;\;.
\ee
For example one can consider the even and odd regularisations
\be
\label{obeta}
\left.
\begin{array}{l}
e^\epsilon_1(t) = (t^2+\epsilon^2)^{(1-a)/2}\\
e^\epsilon_2(t) = t(t^2+\epsilon^2)^{-a/2}
\end{array}
\right\}\quad\Ra\quad \Omega_\epsilon^2(t) = a(1-a) \frac{t^2-\beta_k\epsilon^2}{(t^2 + \epsilon^2)^2}
\quad
\left\{
\begin{array}{l}
\beta_1 = 1/a\\ \beta_2 = 3/(a-1)
\end{array}
\right.
\ee
Note that $\beta_k>0$ for $a>1$ so that these regularised frequencies
display the anticipated behaviour that they change sign for sufficiently
small values of $\eta$. Note also that for $t\ra 0$ these solutions behave
as $e^\epsilon_k(t)\sim 1,t$ respectively, leading to the two alternative
Rosen forms of the (evidently non-singular) Minkowski metric in that
region ($t=0$ being a mere coordinate singularity in the latter case).

Thus these regularised frequencies $\Omega_{\epsilon}^2(t)$ admit
fine-tuned solutions by construction. It is now easy to give more general
multi-parameter families of regularised frequencies that accomplish
this, simply by modifying the regularised solutions $e^\epsilon_k(t)$
by terms that are subleading as $t\ra\infty$ or $\epsilon\ra 0$.

\subsection{Adding the Dilaton}

Plane waves (more generally pp-waves) give rise to
exact string theory backgrounds provided that they are e.g.\ 
supplemented by a null dilaton $\phi=\phi(z^+)$ satisfying the Einstein-dilaton
equation $R_{++}=-2\del_+\del_+ \phi$ where $R_{++}$ is the only non-vanishing component of
the Ricci tensor of the string-frame metric.\footnote{In the Einstein frame, the equation
is $R^e_{++}= (\del_+\phi)^2/2$ so all the backgrounds considered here satisfy the Einstein
equations with a standard, postive-definite, lightcone energy density.}
In the present (scale-invariant and isotropic) case
this equation reduces to
\be
\label{urdil}
\ddot\phi(t) = 4a(a-1)t^{-2} \quad\Ra\quad 
\phi(t) = - 4a(a-1) \log |t|
\ee
where we have selected the solution appropriate to asymptotically 
weak string coupling for $a>1$, and correspondingly a strong string
coupling singularity (the general solution also includes the 
general solution of the homogeneous equation, i.e.\ an affine function
of $t$).

Once we regularise the plane wave profile, we need to
enquire if the regularised dilaton equation 
\be
\label{rdil}
\ddot\phi_\epsilon(t) = -4\Omega_\epsilon^2(t)
\ee
admits solutions with these fine-tuned asymptotics or
if linear terms necessarily arise that give rise to infinite
string coupling in the far past or future. 
It is only in the former case that the pair
$(\Omega_\epsilon^2,\phi_\epsilon)$ can legitimately be considered to
be a regularisation of the original metric-dilaton background.\footnote{The regularisation
of the dilaton was also discussed in \cite{craps2}, but the issues that arise and are relevant
in the present (strong coupling) context are in some sense the opposite of those encountered there.}

That this is indeed an issue can easily be seen explicitly in the toy
model regularisation of \eqref{toy1} or for the regularisations given
in \eqref{obeta}. E.g.\ in the latter case the general solution of
the dilaton equation is 
\be
\phi_\epsilon(t) = 2a(a-1)\left[(1-\beta_k)(t/\epsilon) \arctan(t/\epsilon) - \log
(1+(t/\epsilon)^2)\right] + C(t/\epsilon) + D\;\;.
\ee
For generic values of $\beta_k$ this can be arranged to have the appropriate asymptotics
for either $t\ra+\infty$ or $t\ra -\infty$ by suitable choice of integration
constant $C$ (and with $D=-4a(a-1)\log\epsilon$), but not for both simultaneously. 
This is only possible if $\beta_k=1$
(with the choice $C=0$), and comparison with \eqref{obeta} shows that this corresponds
to the values $\beta_1=1 \ra a=1$ (which is irrelevant) or $\beta_2=1 \ra a=4$. Thus
when one includes the dilaton in the story, the specific regularised profile given
above works only for one specific value of the parameter $a$ (and for other values
other regularisations are required). 

The upshot of this is that the requirement that the regularised profile
admit a fine-tuned dilaton with weak coupling asymptotics in the past
and future gives a further restriction on the regularised profile beyond
those arising from regularity of the GY function (and its fine-tuning
issue). We will return to the issue of how relevant this requirement
actually is in the concluding section 4.

%
%

\subsection{Necessity to go beyond Single Scale Regularisations}

While the above seems encouraging, so far it does not allow us to
claim that there is a well-defined evolution across $t=0$ (here we 
disagree, at least semantically, with \cite{craps2})
and/or that
expectation values of operators are now well-defined at $t=0$.
Indeed, the above analysis, in particular the requirement 
\eqref{Fft} that $\exists \lim_{\epsilon\ra 0} F^\epsilon_{t_i}(t)$
for $t_i<0, t>0$ does not even address the issue what happens at
$t=0$.

Therefore let us for instance look at the $\epsilon\ra 0$ limit of
the GY solution $F_{t_i}^\epsilon(t)$ \eqref{gyeps} when one of the
arguments is $0$, say $t_i=0$, while $t>0$. Since we have chosen
the two solutions $f_k(\eta)$ of the regularised equation of motion
\eqref{Feq} to be $\epsilon$-independent, the $f_k(0)$ are finite and
$\epsilon$-independent as well. One can then immediately read off that
in the limit $\epsilon\ra 0 $ or $t\ra\infty$ the GY function is a
linear combination of terms of the form $\epsilon (t/\epsilon)^a$ and
$\epsilon(t/\epsilon)^{1-a}$ (the addition factor of $\epsilon$ arising,
as before, from the Wronskian). As a consequence of the (single scale)
structure of the regularisation, the divergence structure of the GY
function for $\epsilon\ra 0$ is thus exactly that of the unregularised
GY function \eqref{gyshpw} as $t_i\ra 0$, 
\be
\label{GYregunreg}
\epsilon\ra 0: \quad F^\epsilon_{t_i=0}(t) \sim \epsilon^at^{1-a},\epsilon^{1-a}t^a
\ee
and the $\epsilon\ra 0$ limit of the 
GY function $F^\epsilon_{t_i=0}(t)$ will necessarily 
be either 0 or $\infty$ for all $t>0$.
Analogously, the single scale form of the regularisation implies that
for $t=0$ the divergence structure of the would-be regularised (finite
$\epsilon$) expectation values as $\epsilon\ra 0$ is identical to that
of the expectation values \eqref{abc2} of the unregularised ($\epsilon=0$) theory as $t\ra 0$,
\be
\label{abc3}
\begin{aligned}
\epsilon \ra 0:\quad &\langle \psi|\hat{z}^2|\psi\rangle(t=0)\sim \mathbb{A} \epsilon^{2-2a}+\mathbb{B}
\epsilon^{2a} + \mathbb{C} \epsilon\\
& \langle \psi|\hat{H}_{bc}|\psi\rangle(t=0) \sim \mathbb{A} \epsilon^{-2a}+\mathbb{B} \epsilon^{2a-2} +
\mathbb{C} \epsilon^{-1}\;\;.
\end{aligned}
\ee
Likewise, the classical asymptotics \eqref{urdil} translate into the statement that the 
$\epsilon\ra 0$ behaviour of the regularised dilaton at $t=0$ is given by 
$ \phi_\epsilon(t=0)\sim -4a(a-1)\log |\eps|$.

These facts illustrate that, even at the point particle level, i.e.\
disregarding field-theoretic infinite particle (or rather worlsheet
string mode) production issues, the single scale regularisation advocated
in \cite{craps2} cannot be considered to give rise to a well-defined
evolution from some initial time $t_i<0$ across $t=0$ to some final time
$t_f>0$. Nevertheless, as we will discuss in section 4, these divergences
may appear in a somewhat different light when considered from the point
of view of the space-time from which the plane wave arises as a Penrose limit.

The above considerations do not rule out the possibility that there
are more baroque regularisations, not of the simple single-scale form
\eqref{ssr}, that do give finite $\eps\ra 0$ answers even at $t=0$, and
here we show, by way of example and as a proof of concept, that this
is indeed possible. We will not elaborate on this, however,
since this clearly introduces even more ambiguity into the regularisation
procedure.

One can for instance modify the simple regularisation \eqref{toy1}
by adding a further smaller region around $\eta=0$ with a potential well of 
depth $-\kappa^{2}$, something like 
\be 
\label{toy2}
\Omega^2(\eta) = a(1-a) \eta^{-2} H(|\eta|-\eta_0) + \lambda^2 H(\eta_0 - |\eta|) -(\kappa^2+\lambda^2) 
H(\eta_0/2 - |\eta|)\;\;.
\ee
Analysing the conditions for existence of a GY function that is finite and 
non-zero in the $\epsilon\ra 0$ limit even when one of its arguments is zero, 
one finds that there are non-trivial solutions provided that $\kappa=\kappa(\epsilon)$
depends suitably (and non-trivially) on $\epsilon$ (which necessarily requires the 
introduction of new dimensionful parameters). 

Once one has a well-defined non-singular GY function $F_{t_i}(t)$,
one can use it and its (well-defined and non-singular) dual solution to
determine expectation values from \eqref{abc1}, and these are then also
non-singular as $\epsilon \ra 0$. However, it is not guaranteed (and may
also be too much to hope for) that also the dilaton is then regularised,
in the sense that the limit $\lim_{\epsilon\ra 0}\phi_\epsilon(t=0)$
exists (and this is not true e.g.\ for the above profile \eqref{toy2}
or simple modifications thereof that we have analysed).

\section{Discussion}

Having settled, in section 2, the issue what are the right variables to
use (namely Brinkmann variables), in section 3 we analysed various aspects
of the (matrix) string dynamics in the singular scale-invariant plane
wave backgrounds that arise naturally in this context (e.g.\ from Penrose
limit \cite{bbop1,bbop2} or symmetry \cite{mmpwbb} considerations).

It has become increasingly clear in recent years that a non-perturbative
matrix string description in the spirit of the CSV matrix big bang model
\cite{csv} is not, by itself, sufficient to resolve the null singularity
of the space-time background (cf.\ also the assessment and discussion in
\cite{craps4}). For instance, as argued in \cite{mols}, for the plane wave
matrix big bang models of \cite{mmpwbb} near the singularity the usual
quartic interaction term for the non-Abelian matrix string coordinates
becomes irrelevant, due to the characteristic and inevitable presence
of tachyonic mass terms in these models (for this reason, formally this
argument applies to all the modes of the matrix string, not just the
quantum mechanical zero mode). As a consequence, it does not appear that
(in the spirit of the hope expressed in the context of the original matrix
big bang model of \cite{csv}) the extra non-geometric degrees of freedom
at the singularity arising from the weakly-coupled non-Abelian matrix
string (which are actually also far from massless in the present case)
can lead to a resolution of the singularity in this class of models.
However, it would certainly be desirable to gain a better understanding
of the non-linear dynamics of these models by other means.

Moreover, as far as other degrees of freedom are concerned, the decoupling
arguments of \cite{csv,mmpwbb} for closed and massive open string
degress of freedom appear to be pretty robust.  However, there may be
some room for doubt, as the question of orders of limits becomes somewhat
delicate when dealing simultaneously with Seiberg-Sen, near-singularity,
and regularisation parameter $\epsilon\ra 0$ limits.  There may also be
some subtleties with the DLCQ set-up itself for time-dependent systems
(cf.\ the comments in \cite{craps4}).

In the absence of an intrinsic regularisation mechanism or extra
degrees of freedom in these models, it becomes necessary to choose some
regularisation prescription in order to define the theory. While at this
point any particular choice of prescription is necessarily somewhat ad
hoc, one can nevertheless hope to learn something about the physics
of the singularity by determining what kinds of regularisations are
successful and what features they share.

A very natural first step, and one which we did not question here, is to
regularise the plane wave profile itself, as this at the very least
preserves the $\alpha'$-exact nature of the string background required
for consistency of the matrix string model.  Within this setting, we
investigated the proposal of \cite{craps2} to consider single-scale
regularisations, depending on a single dimensionful regularisation
parameter $\epsilon$, which thus additionally reflect and, as much
as possible, respect, the characteristic scale-invariance of the
unregularised background.

As we have shown in detail, this class of regularisations can at best be
considered to give a formal asymptotic (S-matrix-like) mapping between the
dynamics before and after the singularity, without, however, providing
us with a regularisation of the dynamics (time-evolution of states and
expectation values) through the singularity. We have also traced back
the failure of this class of regularisations to the single-scale nature
of the (attempted) regularisation.

We further found that allowing more complicated, multiple-scale,
regularisations, i.e.\ modifying the profile by terms peaked around
the originally singular locus and introducing additional dimensionful
parameters beyond the overall scale $\epsilon$, say, one can obtain a
non-singular evolution through the singularity. We have not elaborated
nor dwelled upon this, however, because it evidently opens up a lot
of arbitrariness and seems to be indicating that one is missing (rather
than learning) something about the physics of the singularity.

To better understand the physics of what is happening at the singularity
in the models that we have analysed, recall that the plane wave metrics
appearing in the plane wave matrix string models are to be considered
as the lowest-order terms in a (covariant Penrose-Fermi \cite{fermi})
expansion of the original (and singular) space-time metric (or string
background) around its (singular) plane wave Penrose limit. Some of
the divergences that one encounters appear in a different light from
this perspective.

In particular, one characteristic feature of the singularities of the
strong string coupling ($a>1$) backgrounds is that particles are expelled
to $z=\infty$ in finite time (and strings are infinitely stretched, without
actually going through the singularity \cite{hs,hs2}). A regularisation
of this behaviour seems to require something novel to happen, at least some
boundary condition, at $z=\infty$. However, this strongly repulsive behaviour
(which, see footnote 11, is not due to the violation of some energy condition)
is not necessarily something that is exhibited by the original space-time metric
prior to the Penrose limit, but is actually just an artefact of the Penrose limit
truncation of the metric.  

Indeed, the Penrose limit procedure involves first a choice of null-geodesic
and then a suitable infinite rescaling of the metric which has the effect
of infinitely expanding an infinitesimal tubular neighbourhood of this
null geodesic to produce the entire plane wave space-time. Thus,
from this perspective, what happens when particles reach $z=\infty$
in the plane wave is that they reach the boundary of an infinitesimal
neighbourhood of the geodesic in finite time, which should not be a
cause of major concern.\footnote{Note that this interpretation is
not available for the weak string coupling singularities, corresponding
to the parameter range $0<a<1$, since in this case particles and strings
are infinitely squeezed and excited as they approach the singularity.}

This suggests that this particular manifestation of the divergence can be
cured, or at least significantly altered, by retaining higher-order
terms in the Penrose-Fermi expansion. This would also allow one to move
away from null singularities, and this in turn may be good news anyway,
as studying null singularities in a setting in which they are rigidly
null may be reason for concern because of their potential tendency to
deform into spacelike singularities. However, this begs the question if
or how these terms can be incorporated into the matrix string theory,
and we have no answer to this at present.

Similar statements can be made regarding the regularisation of the
dilaton.  We investigated the way that the dilaton depends on the
regularisation, and found that the strongest constraints from dilaton
regularity actually come from the elimination of diverging linear
dilatons at large times $t$. This would be relevant if one thinks of
this as a model for big bang and emergent space-time, in the spirit
of \cite{csv}.  However, in that case one needs to study the late-time
behaviour of the non-Abelian
theory \cite{mols,craps3,MOSinprep}. On the other hand if, in the spirit of the
Penrose limit, one considers the plane wave only as a near-singularity
approximation of the full metric, then the large-$t$ behaviour is
irrelevant simply because it is also an artefact of the approximation.

More generally, and in conclusion, we feel that further work on these
models should focus on (a) alternative methods to analyse and
quantify the behaviour of 1+1 dimensional Yang-Mills theory with vanishing
coupling, (b) perhaps a careful reanalysis of the DLCQ / decoupling arguments 
of \cite{csv,mmpwbb}, and (c) the inclusion of
higher-order terms in the Penrose-Fermi expansion in the matrix string models.

\subsection*{Acknowledgements}

We are grateful to Ben Craps for numerous (and always enjoyable) discussions on these
and related topics over the years. We have also benefitted from discussions with
K.\ Narayan about his work.

This work has been supported by the Swiss National Science Foundation and
the ``Innovations- und Kooperationsprojekt C-13'' of the Schweizerische
Universit\"atskonferenz SUK/CUS.

\appendix

\section{Brinkmann and Rosen Dynamics}

\subsection{Classical Brinkmann and Rosen Dynamics}

The metric of a gravitational plane wave is 
\be
\label{mpwbc}
ds^2= 2dz^+dz^- + A_{ab}(z^+)z^az^b (dz^+)^2 + \d_{ab}dz^a dz^b
\ee
in Brinkmann coordinates $\{z^A\}=\{z^+,z^-,z^a\}$, and 
\be
\label{mpwrc}
ds^2 = 2 du dv + g_{ij}(u) dx^i dx^j
\ee
in Rosen coordinates $\{x^I\}=\{u,v,x^i\}$.
The coordinate transformation 
between the Rosen and Brinkmann forms of a plane 
wave metric has the form
\be
\label{rcbc2}
(u,v,x^i) = (z^+ , 
z^- + \trac{1}{2}\dot{{e}}_{ai}{e}^i_{\;b}z^a z^b, 
{e}^i_{\;a} z^a)\;\;,
\ee
where
$e^i_{\;a}=e^i_{\;a}(u) $ is a vielbein for $g_{ij}(u)$ (satisfying 
the symmetry condition 
$M_{ab} \equiv \dot{{e}}_{ai}{e}^i_{\;b} = \dot{{e}}_{bi}{e}^i_{\;a}$),
and the relation between $g_{ij}(u)$ and $A_{ab}(z^+)$ can be compactly 
written as \cite{gibbonsPW,mmhom}
\be
\label{aee}
A_{ab}(z^+)= \ddot{{e}}_{ai}(z^+) {e}^i_{\;b}(z^+)\;\;.
\ee
Following \cite{gibbonsPW}, it is straightforward to shows that $\mathsf{e}:=\det(e_i^a)$
satisfies 
\be
\ddot{\mathsf{e}}/\mathsf{e} = \Tr A + \left( (\Tr M)^2 - \Tr(M^2)\right) \leq \Tr A = -R_{uu}\;\;,
\ee
with $R_{uu}=R_{++} = -\d^{ab}A_{ab}$ the only non-vanishing component of the Ricci tensor
of a plane wave metric. In particular, if $R_{uu}>0$, then $\mathsf{e}(u_0)=0$ for some finite
value of $u_0$, and the Rosen coordinate system breaks down there.

In the lightcone gauge $z^+=t$ (resp.\ $u=t$), the geodesic Lagrangian 
for a particle in Brinkmann (resp.\ Rosen) coordinates is
\be
\label{lagbcrc}
L_{bc}(z) = \trac{1}{2}(\d_{ab}\dot{z}^a \dot{z}^b +A_{ab}(t)z^a z^b)
\quad,\quad
L_{rc}(x) = \trac{1}{2}g_{ij}(t)\dot{x}^i \dot{x}^j\;\;.
\ee
These two Lagrangians are equivalent in the sense that 
under the time-dependent coordinate transformation $x^i=e^i_{\;a}(t) z^a$ 
(the essential part of the transformation between Brinkmann and Rosen coordinates) 
they transform into each other up to a total time-derivative.

In the body of the paper we will focus on isotropic plane waves,
characterised by $g_{ij}(t)=e(t)^2\delta_{ij}$.
In that case, the
Hamiltonian descriptions \eqref{ihbcrc} of the systems described by the Lagrangians
\eqref{ilbcrc} are related by the linear time-dependent canonical
transformation
\be
\label{ct}
(x,p_x)=(z/e(t),e p_z - \dot{e}z) \quad\LRa\quad (z,p_z) = (e(t)x,p_x/e(t) + x \dot{e}(t)) 
\ee
that can be described by a generating function of the $2^{\mathrm{nd}}$ kind,
\be
\label{gefcantra}
F_2(z, p_x;t)=p_x \frac{z}{e(t)}+\frac{1}{2}z^2\frac{\dot{e}(t)}{e(t)}:\quad
\left\{\begin{array}{l}
x=\frac{\partial F_2}{\partial p_x}, p_z= \frac{\partial F_2}{\partial z}\\
H_{rc}=H_{bc}+\frac{\partial F_2}{\partial t}\;\;.
\end{array}\right.
\ee
Given a specific Rosen metric, specified by a solution $e(t)$ to the harmonic oscillator equation, 
one can construct its linearly independent
Wronskian partner $\tilde{e}(t)$, 
\be
\label{etilde}
\tilde{e}(t) = e(t) \int^t_{t_0}\frac{\der t'}{e(t')^2}\equiv e(t) T(t):\quad
W(e,\tilde{e})\equiv e\dot{\tilde{e}}-\dot{e}\tilde{e}=1\;\;,
\ee
and consider the corresponding ``dual'' Rosen system, arising from the dual Rosen metric with
$\tilde{g}_{ij}(t)=\tilde{e}(t)^2\d_{ij}$.
In the classical theory, 
the geometric significance of the function $T(t)$ introduced above 
is that of \textit{conformal time} for the
Rosen metric (isotropic plane waves are conformally flat)
\be
ds^2 = 2dtdx^- + e(t)^2 dx^2 = e(t)^2 (2dTdx^- + dx^2)\;\;.
\ee
Its related significance in the quantum Rosen theory is explained in section 2.2.

\subsection{Quantum Brinkmann and Rosen Dynamics}

Since classically the Brinkmann and Rosen systems are related by linear (albeit time-dependent) canonical
point transformations, thus with a generating function that is at most quadratic in the 
canonical variables, it folllows on general grounds that these canonical transformations
can be unambiguously implemented by a unitary (isometric) transformation in the quantum theory.

In the case at hand, this transformation takes the form
\be
\label{upsi}
\begin{aligned}
&\psi_t(z)=\left(U\psi'_t\right)(z)=\frac{1}{\sqrt{e(t)}}\ex{-\frac{\dot{e}(t)}{2
e(t)}iz^2}\psi_t'(z/e(t))\\
&\psi'_t(x)=\left(U^{-1}\psi_t\right)(x)=\sqrt{e(t)}\ex{-\frac{1}{2}i\dot{e}(t)
e(t)x^2}\psi_t(e(t)x)\;\;.
\end{aligned}
\ee
Here the change of argument of the wave function is just (the crucial part of) the 
coordinate transformation between Brinkmann and Rosen coordinates, the phase factor
$\exp -ik(z,t)$ is due
to the fact that the two Lagrangians differ by a total time-derivative, and the prefactor
$\sqrt{e(t)}$ reflects the fact that Brinkmann (resp.\ Rosen) states are to be normalised
with respect to the measure $dz$ (resp.\ $dx=dz/e(t)$). 

This gives an isometry $U: \mathcal{H}_{rc}\ra\mathcal{H}_{bc}$ of the
Rosen and Brinkmann spaces of states (with measures $dx$ and $dz$ respectively)
which implements the canonical transformation
\eqref{ct} in the form
\be
\label{qct}
U^{-1}\hat{z}U=e\hat{x}\qquad
U^{-1}\hat{p}_z U=\dot{e}\hat{x}+\frac{\hat{p}_x}{e}
\ee
and intertwines the action of the Schr\"odinger operators $i\del_t-\hat{H}$,
\be
\label{sbcsrc}
\left(i\del_t-\hat{H}_{bc}\right)\psi_t(z) =
\frac{1}{\sqrt{e(t)}}\;\ex{-ik(z,t)}(i\del_t - \hat{H}_{rc}) \psi_{t}'(x)\;\;.
\ee
To illustrate the mapping \eqref{upsi} between Rosen and Brinkmann states, 
we remark  that the Brinkmann propagator (WKB) state 
\eqref{ptzz} remains a solution of the Schr\"odinger equation if one 
replaces $(F,-G)$ by any pair of solutions $(e(t),\tilde{e}(t)= e(t)T(t))$ with 
Wronskian $+1$, and that this more general solution
\be
\label{psie}
\psi_{t,z_i}(z) = 
\frac{1}{\sqrt{e(t)}}\ex{(i/2)[(\dot{e(t)}/e(t))z^2 - T(t)z_i^2 - 2zz_i/e(t)]}\;\;,
\ee
is precisely the one-parameter family of states
one obtains by mapping the Rosen momentum modes $\psi'_{t,k}(x)$ 
\eqref{ptkx} with $k=-z_i$ to Brinkmann coordinates using \eqref{upsi}.
Conversely, the Rosen-image of the propagator state \eqref{ptzz} under $U^{-1}$ 
is a momentum state of that Rosen metric whose metric is given by (the square of)
the GY function \eqref{gye},  $e(t)=F_{t_i}(t)$.

In order to verify \eqref{sbcsrc} it is useful to note that 
it follows from \eqref{qct} that for any (reasonable, polynomial say)
Brinkmann operator $\hat{O}_{bc}=\hat{O}(\hat{z},\hat{p}_z)$ one has
\be
U^{-1}\hat{O}_{bc}U =\hat{O}(U^{-1}\hat{z}U,U^{-1}\hat{p}_zU) =
\hat{O}(e\hat{x},\dot{e}\hat{x}+ (1/e)\hat{p}_x) \equiv (\hat{O}_{bc})_{rc}\;\;,
\ee
i.e.\ one obtains the Brinkmann operator expressed in terms of Rosen variables. 
In particular, 
at the level of expectation values, and for Brinkmann and Rosen states related by $\psi=U\psi'$ 
one generally has
\be
\label{obc}
<\psi|\hat{O}_{bc}|\psi> = <\psi'|(\hat{O}_{bc})_{rc}|\psi'>\;\;,
\ee
while $U$ also implements the classical relation
$H_{bc}-H_{rc}=-\partial_t{F_2}$ 
\eqref{gefcantra}
at the operator level,
\be
\label{qhh}
\begin{aligned}
U^{-1}\hat{H}_{bc} U &= (\hat{H}_{bc})_{rc}  &=
\hat{H}_{rc} - (\hat{\dot{F}}_2)_{rc} \\
U\hat{H}_{bc} U^{-1} &= (\hat{H}_{rc})_{bc} 
&= \hat{H}_{bc} + (\hat{\dot{F}}_2)_{bc}\;\;,
\end{aligned}
\ee
and at the level of expectation values, 
\be
\label{hvev}
 <\psi|\hat{H}_{bc}|\psi> -<\psi'|\hat{H}_{rc}|\psi'> 
= - <\psi'|(\hat{\dot{F}}_2)_{rc}|\psi'> = - <\psi|(\hat{\dot{F}}_2)_{bc}|\psi>\;\;,
\ee
where $(\hat{\dot{F}}_2)_{bc/rc}$ are the symmetrically ordered operators
\be
\label{Fhat}
\begin{aligned}
(\hat{\dot{F}}_2)_{bc}&= 
-\trac{1}{2}(\dot{e}/e) (\hat{p}_z \hat{z} + \hat{z} \hat{p}_z) +\trac{1}{2}(\dot{e}^2 - \omega^2
e^2)\hat{z}^2/e^2 \\
(\hat{\dot{F}}_2)_{rc}&= 
-\trac{1}{2}(\dot{e}/e)(\hat{p}_x \hat{x} + \hat{x} \hat{p}_x) 
- \trac{1}{2}(\dot{e}^2 + \omega^2 e^2) \hat{x}^2 \;\;.
\end{aligned}
\ee
In order to determine these expectation values for the Rosen wave packets \eqref{rwp},
we need to calculate the expectation values of $\hat{p}_x\hat{x}+\hat{x}\hat{p}_x$ and $\hat{x}^2$, the
latter being of independent interest anyway. Straightforward calculations lead to
\be
\langle \psi'|(\hat{p}_x\hat{x}+\hat{x}\hat{p}_x)|\psi'\rangle =\mathbb{C}+2 T(t) \mathbb{A}
\ee
and
\be
\label{expvx}
\langle \psi'|\hat{x}^2|\psi'\rangle = \mathbb{A}+ T(t) \mathbb{C} + T^2(t) \mathbb{B}\;\;,
\ee
where
\be
\label{ABC}
\mathbb{A}= \int\der k\,|a'(k)|^2\qquad
\mathbb{B}=\int\der k\,k^2|a(k)|^2\qquad
\mathbb{C}=\int\der k\, ik\left(a(k)^*a'(k)-a'(k)^*a(k)\right)\;\;,
\ee
and we note for a nontrivial wave packet the terms $\mathbb{A}$ and $\mathbb{B}$
will be non-zero while $\mathbb{C}$ can be zero (e.g.\ if all the $a(\ell)$ are real). 
The 
expectation value of $\hat{z}^2$ can then be deduced from \eqref{z2} and \eqref{expvx}, 
\be
\label{expvz}
\langle\psi |\hat{z}^2 |\psi\rangle = e(t)^2 \mathbb{A} + \tilde{e}(t)^2 \mathbb{B} +
e(t)\tilde{e}(t)\mathbb{C}\;\;,
\ee
while from \eqref{hrcvev} and 
\eqref{hvev}
one finds for the expectation value of the Brinkmann Hamiltonian
\be
\label{expvalhbc}
\langle\psi|\hat{H}_{bc}|\psi\rangle=
\frac{1}{2}(\dot{e}^2+\omega^2e^2)\mathbb{A}+\frac{1}{2}(\dot{\tilde{e}}^2
+\omega^2\tilde{e}^2)\mathbb{B}+\frac{1}{2}(\dot{e}\dot{\tilde{e}}+\omega^2e\tilde{e})
\mathbb{C}\;\;.
\ee
Here we have made use of the fact that the dual solution $\tilde{e}(t)$ \eqref{etilde}
satisfies
\be
\tilde{e}(t)=e(t)T(t)\quad\Ra\quad \dot{\tilde{e}}(t)= \dot{e}(t)T(t) + e(t)^{-1} \;\;,
\ee
and introduced the functions
$E(e)=  \frac{1}{2}(\dot{e}^2+\omega^2e^2)$,$E(\tilde{e})=
\frac{1}{2}(\dot{\tilde{e}}^2+\omega^2\tilde{e}^2)$ and
$E(e, \tilde{e})= 1/2(\dot{e}\dot{\tilde{e}}+\omega^2e\tilde{e})$.

\rnc{\Large}{\normalsize}

\end{document}